\documentclass[preprint,aps,pra,showpacs,floatfix]{revtex4}
\usepackage{amsmath}
\usepackage{times}
\usepackage{euscript}
\usepackage{graphicx}
\usepackage{amsthm}
\usepackage{epsf}
\usepackage{epsfig}
\usepackage{bm}
\graphicspath{ {./Figures/} }
\renewcommand{\vec}[1]{{\mbox{\boldmath$#1$}}}

\begin{document}
\thispagestyle{empty}
\title{Electron-positron pair creation in low-energy 
collisions of heavy bare nuclei}
\author{I.~A.~Maltsev,$^{1, 2}$
V.~M.~Shabaev,$^{1}$
I.~I.~Tupitsyn,$^{1}$
A.~I.~Bondarev,$^{1}$
Y.~S.~Kozhedub,$^{1,3}$
G.~Plunien,$^{4}$
and Th.~St\"ohlker$^{5,6,7}$
}
\affiliation{
$^1$ Department of Physics, St. Petersburg State University,
Ulianovskaya 1, Petrodvorets, 198504 St. Petersburg, Russia\\
$^2$ ITMO University, 
Kronverkskii ave 49, 197101 St. Petersburg, Russia\\
$^3$ SSC RF ITEP of NRC ”Kurchatov Institute”,
Bolshaya Cheremushkinskaya 25,
117218 Moscow, Russia\\
$^4$ Institut f\"ur Theoretische Physik, Technische Universit\"at Dresden,
Mommsenstra{\ss}e 13, D-01062 Dresden, Germany\\
$^5$
GSI Helmholtzzentrum f\"ur Schwerionenforschung GmbH,
Planckstrasse 1, D-64291 Darmstadt, Germany \\
$^6$Helmholtz-Institute Jena, D-07743 Jena, Germany\\
$^7$Institut f\"ur Optik und Quantenelektronik,
Friedrich-Schiller-Universit\"at,
D-07743 Jena, Germany
\vspace{10mm}
}
%
\begin{abstract}
A method for calculations of electron-positron pair-creation 
probabilities in low-energy heavy-ion collisions is developed. 
The approach is based on the propagation of all one-electron states via 
the numerical solving of the time-dependent Dirac equation 
in the monopole approximation. The electron wave functions are 
represented as finite sums of basis functions constructed from 
B-splines using the dual-kinetic-balance technique. The calculations 
of the created particle numbers and the positron energy spectra
are performed for the collisions of bare nuclei at the energies near 
the Coulomb barrier with the Rutherford trajectory 
and for different values of the nuclear charge 
and the impact parameter. To examine the role of the spontaneous 
pair creation the collisions with
a modified velocity and with a time delay are also considered. 
The obtained results are compared with the previous calculations
and the possibility of observation of the spontaneous pair creation is 
discussed.
\end{abstract}
\pacs{ 34.90.+q, 12.20.Ds}
\maketitle
\section{INTRODUCTION}
The stationary Dirac equation leads to a singularity in the 
solution for the ground state of an electron in 
the field of the pointlike nucleus with the charge $Z>$137. 
But for an extended nucleus the energy of the 1$s_{1/2}$~state $E(Z)$ 
goes continuously beyond the point $Z=137$ and reaches the 
negative-energy continuum 
at the critical value $Z_{\rm C}\approx173$~\cite{Pomeranchuk_45, Gershtein_70, 
Greiner_69, Zeldovich_71}. 
As predicted independently by Gershtein and Zeldovich~\cite{Gershtein_70} and 
by Pieper and Greiner~\cite{Greiner_69}, if the empty level dives into 
the negative-energy continuum, then it turns into a resonance that
can lead to the spontaneous decay of the vacuum via emission of 
a positron and occupation of the supercritical K-shell by an electron.
The experimental observation of this effect would confirm
the predictions of quantum electrodynamics in the highly nonperturbative 
supercritical domain. Unfortunately, the charge of the heaviest produced nuclei 
is far less than the required one, $Z_{\rm C}$. However, in the collision of two ions, if
their total charge is sufficiently large, the ground state of the formed quasimolecular
system can become so deeply bound that the spontaneous pair creation is possible. 
The most favorable collision energy for investigation of the supercritical regime 
is about the Coulomb barrier~\cite{Greiner_85}.
In heavy-ion collisions, the electron-positron pairs can also be created dynamically due 
to the time-dependent potential of the moving ions. In order to find the signal from the vacuum
decay one needs to distinguish the spontaneously produced pairs from the dynamical background.

The experiments for searching the spontaneous pair creation were performed at 
GSI (Darmstadt, Germany)
using the collisions of partially stripped ions with neutral atoms, 
but no evidence of the vacuum decay was found~\cite{Muller_94}. 
It should be noticed that for studying this phenomenon the collisions of bare nuclei
would be more favorable due to the empty K-shell.
It is expected that the upcoming Facility for Antiproton and Ion Research~(FAIR)
will provide 
new opportunities for investigations of low-energy heavy-ion collisions, 
probably including the collisions of fully stripped ions~\cite{fair_design_01, Gumbaridze_09}.

To date a number of approaches to calculations of various processes 
in low-energy heavy-ion collisions have been 
proposed~\cite{Gershtein_73, Popov_73, Peitz_73, Reinhardt_81, Muller_88, Kurpick_92, 
Ackad_07, Ackad_08, Tup_10, Tup_12, Kozhedub_2013, 
Bondarev_2013, Maltsev_2013, Mac_12, Deyneka_13, Kozhedub_14, Khriplovich_14}.
In Refs.~\cite{Gershtein_73, Popov_73, Peitz_73},
the pair-creation process was considered in the static approximation,
according to which the corresponding probability is proportional to the resonance width 
$\Gamma (R)$ which depends on the internuclear distance $R(t)$. Such an approximation does not 
take into account the dynamical effects. A more advanced approach was based on the propagation
of a finite number of initial states using the time-dependent adiabatic basis set 
with the Feshbach projection 
technique~(see Refs.~\cite{Reinhardt_81, Greiner_85, Muller_88} and references therein). 
This method allowed calculations 
of the pair-creation probabilities
employing small numbers of the basis functions. However,
the small basis size might lead to the low resolution of the continuum.
From the results, which were basically obtained in the monopole approximation, 
it was concluded that the spontaneous contribution 
is indistinguishable from the dynamical background in the positron spectra in 
elastic collisions, and only in hypothetical collisions with the nuclear sticking 
can there be the visible effects of the vacuum decay~\cite{Reinhardt_81, Muller_88}.
Another dynamical approach~\cite{Ackad_07, Ackad_08} 
was based on solving the time-dependent
Dirac equation in the monopole approximation with the mapped Fourier grid method. 
In Ref.~\cite{Ackad_08}, the pair-creation probabilities 
were calculated with propagation of all initial states of
a very large basis set, compared to the previous works, 
that might improve the energy resolution of the continuum.
For the collisions of bare uranium nuclei, the results for the positron spectra
were quite different from those in Ref.~\cite{Muller_88}.
The importance of the dynamical pair-production effects follows also from the
recent perturbative evaluation of Ref.~\cite{Khriplovich_14}.

In the present work, we develop an alternative method for calculations of the
pair-creation probabilities in low-energy heavy-ion collisions. In this method,
the time-dependent Dirac equation is solved numerically in the monopole approximation 
employing the stationary basis set.
The basis functions are constructed from the B-splines using 
the dual-kinetic-balance (DKB) approach~\cite{Shabaev:prl:04},
which prevents the appearance of nonphysical spurious states. 
The DKB B-spline basis set provides
a very accurate representation of the continuum and previously 
was successfully used in QED calculations
for the summation over the whole Dirac spectrum 
(see, e.g., Refs.~\cite{Shabaev_05, Kozhedub_10}).
All of the eigenvectors of the initial Hamiltonian matrix are propagated in order to obtain
the one-electron transition amplitudes, which are used to calculate the particle-production 
probabilities. The calculations are performed for the symmetric 
collisions of bare nuclei with different values of the nuclear charge at the energy near
the Coulomb barrier.

The paper is organized as follows. The pair-creation process in a time-dependent 
external field is briefly discussed in 
Sec.~\ref{subsec:pair_creation}. The monopole approximation 
is considered in Sec.~\ref{subsec:monopole}.
The method for solving the time-dependent Dirac equation is described 
in Sec.~\ref{subsec:finite_dirac}.
The obtained results and their comparison with the previous calculations are 
presented in Sec.~\ref{sec:results}.

The relativistic units ($\hbar=c=1$) are used throughout the paper.
\section{THEORY}
\label{sec:theory}
\subsection{Pair creation in external field}
\label{subsec:pair_creation}
In the present work we take into account the interaction of electrons with
the strong external field nonperturbatively, 
but neglect the electron-electron interaction,
assuming the electrons can influence each other only via the Pauli exclusion 
principle. 
The positron states as well as the creation of electron-positron pairs can
be treated within the Dirac original model where the negative-energy states
are considered to be initially occupied by electrons. The production of an
electron-positron pair appears as a transition of an electron from 
the negative-energy continuum to a positive-energy state, in formal
agreement with quantum electrodynamics. The negative-energy electron 
states properly transformed describe the states of positrons,  which
in the mentioned model correspond to the holes in the filled lower 
continuum.

The one-electron dynamics is determined by the time-dependent Dirac equation
\begin{equation}
 i \frac{\partial \psi (\vec{r}, t)}{\partial t} = 
 \hat{H}_D (t) \, \psi (\vec{r}, t) \, ,
 \label{eq:time_dirac}
\end{equation}
where
\begin{equation}
\hat{H}_D (t)= \vec{\alpha} \left( \vec{p} - e\vec{A}(t)\right)+V(t)+m_e \beta
\end{equation}
and the potential $(V(t), \vec{A}(t))$ describes the 
interaction with the external field. One can define the 
solutions $\psi^{(+)}_i(\vec{r}, t)$ and $\psi^{(-)}_i (\vec{r}, t)$ of 
Eq.~(\ref{eq:time_dirac}) with the following boundary conditions
\begin{equation}
 \psi^{(+)}_i(\vec{r}, t_{\rm in})=\varphi^{\rm in}_i (\vec{r}), \,
 \psi^{(-)}_i(\vec{r}, t_{\rm out})=\varphi^{\rm out}_i (\vec{r}),
\end{equation}
\begin{equation}
 \hat{H}_D(t_{\rm in}) \, \varphi^{\rm in}_i (\vec{r})=\varepsilon^{\rm in}_i \varphi^{\rm in}_i (\vec{r}), \,
  \hat{H}_D(t_{\rm out}) \, \varphi^{\rm out}_i (\vec{r})=\varepsilon^{\rm out}_i \varphi^{\rm out}_i (\vec{r}),
\end{equation}
where $t_{\rm in}$ is the initial and $t_{\rm out}$ is the final time moment. 
In the final expressions it will be assumed that 
$t_{\rm in} \rightarrow -\infty$ and $t_{\rm out} \rightarrow \infty$.

The formulas for the probabilities 
of pair creation can be derived using 
the second quantization technique~\cite{Greiner_85, Fradkin_91}.
In the Heisenberg picture, one can introduce the time-dependent
field operator $\hat{\Psi} (\vec{r}, t)$ and the time-independent state vectors 
$\arrowvert 0,\rm in\rangle$ and $\arrowvert 0,\rm out\rangle$,
which correspond to the fully occupied negative-energy continua at 
$t_{\rm in}$ and $t_{\rm out}$, respectively. Using the functions 
$\psi_i^{(+)}(\vec{r},t)$ and $\psi_i^{(-)}(\vec{r}, t)$, 
the operator $\hat{\Psi}(\vec{r}, t)$ can be expanded as
\begin{equation}
 \hat{\Psi}(\vec{r}, t)=\sum_{i>F} \hat{b}^{\rm (in)}_i \, \psi^{(+)}_i(\vec{r}, t)+
 \sum_{i<F} \hat{d}^{\rm \, (in) \, \dagger}_i \, \psi^{(+)}_i(\vec{r}, t),
 \label{eq:f_operator1}
\end{equation}
\begin{equation}
 \hat{\Psi}(\vec{r}, t)=\sum_{i>F} \hat{b}^{\rm (out)}_i \, \psi^{(-)}_i(\vec{r}, t)+
 \sum_{i<F} \hat{d}^{\rm \, (out) \, \dagger}_i \, \psi^{(-)}_i(\vec{r}, t).
 \label{eq:f_operator2}
\end{equation}
Here the Fermi level $F$ is the border between the filled 
negative-energy states and the vacant positive-energy states 
($\varepsilon_F=-m_e$),
$\hat{b}^{\rm (in)}_i$ and $\hat{b}^{\rm (out)}_i$
are the annihilation operators for electrons,  and
$\hat{d}^{\rm \, (in) \, \dagger}_i$ and
$\hat{d}^{\rm \, (out) \, \dagger}_i$ are the creation operators for holes (positrons).
They obey the standard anticommutation relations and their action on the vacuum
states is 
\begin{equation}
  \hat{b}_i^{\rm (in)} \arrowvert 0, {\rm in} \rangle = 0, \quad 
  \hat{b}_i^{\rm (out)} \arrowvert 0, {\rm out} \rangle = 0 \quad 
  {\rm for} \, \, i>F
 \label{eq:b_vacuum}
\end{equation}
 and
\begin{equation}
  \hat{d}_i^{\rm \, (in)} \arrowvert 0, {\rm in} \rangle = 0, \quad 
  \hat{d}_i^{\rm \, (out)} \arrowvert 0, {\rm out} \rangle = 0 \quad
  {\rm for} \, \, i<F.
 \label{eq:b_vacuum}
\end{equation}
It should be noted that these operators refer to the physical particles
only at the corresponding time moments $t_{\rm in}$ and $t_{\rm out}$. 

Since we assume that at the initial time moment $t_{\rm in}$, the negative-energy 
continuum is occupied and all of the positive-energy states are free, 
the system is described by the vector $\arrowvert 0, \rm{in} \rangle$. The operators should
correspond to the particles produced at $t_{\rm out}$, which is the measurement time.
By employing Eqs.~(\ref{eq:f_operator1}), and  (\ref{eq:f_operator2}), and the anticommutation relations 
between the annihilation and creation operators, one can derive the following expressions for
the numbers of electrons $n_k$ and positrons $\overline{n}_p$ created in the states  
$k>F$ and $p<F$, respectively~\cite{Greiner_85, Fradkin_91}:
\begin{equation}
  n_k= \langle 0, {\rm in} \lvert \hat{b}^{\rm (out) \, \dagger}_k \hat{b}^{\rm (out)}_k 
  \rvert 0, {\rm in} \rangle = 
  \sum_{i<F} \arrowvert a_{ik} \arrowvert^2 \, ,
 \label{eq:electrons}
\end{equation}
\begin{equation}
  \overline{n}_p= \langle 0, {\rm in} \lvert \hat{d}^{\rm \, (out) \, \dagger}_p 
  \hat{d}^{\rm \, (out)}_p \rvert 0 , {\rm in}\rangle=
  \sum_{i>F} \arrowvert a_{ip} \arrowvert^2 \, ,
 \label{eq:positrons}
\end{equation}
where  
\begin{equation}
 a_{ij}(t)=\int d^3 \vec{r} \, \psi^{(-)\dagger}_i(\vec{r}, t) \, \psi^{(+)}_j (\vec{r}, t)=
 a_{ij}
 \label{eq:tr_ampl}
\end{equation}
are the one-electron transition amplitudes, which are time independent 
because the functions $\psi_i^{(+)}(\vec{r}, t)$ and $\psi_i^{(-)}(\vec{r}, t)$
are solutions of Eq.~(\ref{eq:time_dirac}).
For calculation of $n_k$ and $\overline{n}_p$,
we use the finite-basis-set method and, therefore, in Eqs.~(\ref{eq:electrons})
and~(\ref{eq:positrons}) the summation runs over a finite number
of states. In order to obtain $a_{ij}$,   
the eigenstates $\varphi^{\rm in}_i (\vec{r})$ 
of the initial Hamiltonian $\hat{H} (t_{\rm in})$,
including the bound states and the states of the both discretized
continua, are evolved to the time $t_{\rm out}$ via solving 
the time-dependent Dirac equation and are 
then projected on the eigenstates $\varphi^{\rm out}_j (\vec{r})$ 
of the final Hamiltonian $\hat{H} (t_{\rm out})$: 
\begin{equation}
a_{ij}=\int d^3 \vec{r} \, \varphi^{(\rm out) \dagger}_i(\vec{r}) \, 
\psi^{(+)}_j(\vec{r}, t_{\rm out}).
\label{eq:tout_tr_ampl} 
\end{equation}
The total number of created particles is given by
\begin{equation}
  P=\sum_{k>F} \, n_k = \sum_{p<F} \, \overline{n}_p. 
\end{equation}
In the discrete basis set, the positron energy spectrum can be calculated using
the  Stieltjes method~\cite{Langhoff_74}:
\begin{equation}
 \frac{dP}{dE} \left(\frac{\varepsilon^{\rm out}_p+ \varepsilon^{\rm out}_{p+1}}{2} \right)=
 \frac{1}{2} \frac{\overline{n}_{p+1}+\overline{n}_{p}}{\varepsilon^{\rm out}_{p+1}-\varepsilon^{\rm out}_p}.
\end{equation}

\subsection{Monopole approximation}
\label{subsec:monopole}
We consider the low-energy collision of two heavy bare 
nuclei $A$ and $B$ which move along the classical trajectories.
In the field of the nuclei, the electron dynamics is described by Eq.~(\ref{eq:time_dirac})
with the two-center potential,
\begin{equation}
 V(\vec{r}, t)=V^A_{\rm nucl} \left(\vec{r}-\vec{R}_A (t) \right)
 +V^B_{\rm nucl} \left( \vec{r}-\vec{R}_B (t) \right) \, ,
 \label{eq:two_center} 
\end{equation}
where $\vec{R}_A$ and $\vec{R}_B$ denote the nuclear positions and 
\begin{equation}
 V_{\rm nucl} (\vec{r})=\int d^3 \vec{r}'
  \frac{\rho_{\rm nucl} \left( \vec{r}' \right)}{\lvert \vec{r} - \vec{r}' \rvert}.
\end{equation}
In this paper, we use the uniformly charged sphere model for the nuclear charge-density 
distribution $\rho_{\rm nucl} (\vec{r})$. The vector potential $\vec{A}$ can be neglected 
due to the low collision energy.

The numerical solving of the time-dependent Dirac equation with the two-center
potential~(\ref{eq:two_center}) requires very demanding three-dimensional 
calculations. One may expect, however, that the main contribution to the
pair creation results from the short internuclear distances, where the symmetric
quasimolecular system is well described within the monopole 
approximation~\cite{Reinhardt_81}.
In this approximation, only the  spherically symmetric part 
of the partial expansion
of the potential~(\ref{eq:two_center}) is taken into account:
\begin{equation}
 V_{\rm mon}(r, t)=\frac{1}{4 \pi} \int d \Omega \, V(\vec{r}, t) \, .
 \label{eq:monopole_potential}
\end{equation}
Here we assume that the origin of the coordinate frame is chosen at the center of mass.
For the central field~(\ref{eq:monopole_potential}) the Dirac wave function
can be written as
\begin{equation}
 \psi_{\kappa m}(\vec{r},t) =
  \left  ( \begin{array}{l} \displaystyle
  \,\, \frac{~G_{\kappa}(r,t)}{r} \,  \chi_{\kappa m}(\Omega)
  \\[4mm] \displaystyle
  i \, \frac{F_{\kappa}(r,t)}{r} \, \chi_{-\kappa m}(\Omega)
  \end{array} \right ) \,,
  \label{eq:bispinor}
\end{equation}
where $G_{\kappa}(r,t)$ and $F_{\kappa}(r,t)$ are the large and small radial
components, respectively, $\chi_{\kappa m}(\Omega)$  is the spherical spinor,
and $\kappa=(-1)^{j+l+1/2}(j+1/2)$ is the relativistic angular quantum number. 
Substituting the expression~(\ref{eq:bispinor}) into 
the Dirac Eq.~(\ref{eq:time_dirac}) leads to
\begin{equation}
 i\frac{\partial}{\partial t} \phi (r, t) =\hat{H} (t) \, \phi (r, t) \, ,
 \label{eq:radial_dirac}
\end{equation}
where 
\begin{equation}
 \phi (r, t)= \left(
 \begin{array}{lll}
  G (r, t)\\
  F (r, t)
 \end{array}
 \right)
 \label{eq:two_comp}
\end{equation}
and
\begin{equation}
 \hat{H} (t) = \left( \begin{array}{cc} \displaystyle
  V_{\rm mon}(r,t)+m_e & \displaystyle
  -\frac{d}{dr}+\frac{\kappa}{r} 
  \\[4mm] \displaystyle
  \frac{d}{dr}+\frac{\kappa}{r} & \displaystyle
  V_{\rm mon}(r,t) -m_e \,
  \end{array} \right)
  \label{eq:radial_hamiltonian}
\end{equation}
is the radial Dirac Hamiltonian.

For large nuclear separation, the one-electron energy levels 
of the monopole Hamiltonian $\hat{H}(t)$ are quite different from
the real two-center ones. However, the vacuum state
defined with respect to the instantaneous monopole Hamiltonian at 
some large internuclear distance can be considered as the 
initial state of the system since, as assumed above, 
the particles are mainly produced at short internuclear distances, 
where the monopole approximation is valid. 
The limitation of the employed model is that we can not
isolate the final population of a particular one-electron 
state belonging to one of the nuclei.
\subsection{Dirac equation in a finite basis set}
\label{subsec:finite_dirac}
For solving Eq.~(\ref{eq:radial_dirac}), we employ the time-independent 
finite basis set $\{u_k (r)\}$:
\begin{equation}
 \phi(r, t)= C_{k} (t) \, u_{k} (r) \, ,
 \label{eq:fin_expansion}
\end{equation}
\begin{equation}
 i S_{jk} \, \frac{d C_k (t)}{dt}=H_{jk}(t) C_k (t) \, ,
 \label{eq:fin_dirac}
\end{equation}
where $S_{jk}=\langle u_j \arrowvert u_k \rangle$, 
$H_{jk} (t)=\langle u_j \arrowvert \hat{H}(t) \arrowvert u_k \rangle$, and
the Hamiltonian $\hat{H}(t)$ is defined by Eq.~(\ref{eq:radial_hamiltonian}).
Here and below, the summation over the repeated indices is implied. 
Equation~(\ref{eq:fin_dirac}) is solved using the Crank-Nicolson method~\cite{Crank_47}.
According to this method, for a
short time interval $\Delta t$,
the coefficients $C_{k} (t+\Delta t)$ can be found 
from the system of linear equations
\begin{equation}
 \left [ S_{jk} \, + \frac{i \Delta t}{2} \,\,  H_{jk}(t+\Delta t/2) \, \right ] \, 
 C_k(t+\Delta t) \,=\,
 \left[ S_{jk} - \frac{i \Delta t}{2}\, H_{jk}(t+\Delta t/2) \right ] \, C_k(t).
 \label{eq:lin_system}
\end{equation}
We solve the system~(\ref{eq:lin_system}) employing the iterative BiCGS (BiConjugate Gradient Squared)
algorithm~\cite{Joly_93}. It should be noted that the Crank-Nicolson method conserves the
norm of the wave function at each time step~\cite{Deyneka_13}.

In order to obtain the initial states, one can start from the variational principle
\begin{equation}
  \delta \mathcal{F}=0 \, ,
  \label{eq:var_principle}
\end{equation}
\begin{equation}
 \mathcal{F}=\langle \phi \lvert ( \hat{H}_0-\varepsilon ) \rvert \phi  \rangle \, ,
\end{equation}
which is equivalent to the stationary Dirac equation. The Lagrange multiplier
$\varepsilon$ corresponds to the energy of an eigenstate 
of the instantaneous Hamiltonian $\hat{H}_0=\hat{H}(t_{\rm in})$ at the initial time 
moment $t_{\rm in}$. Substituting the expansion~(\ref{eq:fin_expansion}) into
Eq.~(\ref{eq:var_principle}), one gets the system of
equations
\begin{equation}
  \frac{d\mathcal{F}}{dC_k}=0 \, .
  \label{eq:var_system}
\end{equation}
This system leads to the generalized eigenvalue problem
\begin{equation}
 H_{jk} C_k=\varepsilon S_{jk} C_k \, ,
 \label{eq:gen_eigen}
\end{equation}
which can be solved using the standard numerical routines.

A disadvantage of the straightforward implementation of 
the finite-basis-set method is the presence of nonphysical 
spurious states for $\kappa>0$. To avoid such states,
we employ the DKB approach~\cite{Shabaev:prl:04}. 
According to this approach, the basis functions are constructed as
\begin{equation}
  u_k(r)=\left(
   \begin{array}{c} \displaystyle
   \pi_k (r)
   \\[4mm] \displaystyle
   \frac{1}{2m_e} \left( \frac{d}{dr} + \frac{\kappa}{r} \right) \pi_k (r)
   \end{array}
   \right), \qquad k \leq n \, ,
 \label{eq:dkb_basis1}
\end{equation}
\begin{equation}
 u_{k}(r)=\left(
 \begin{array}{c} \displaystyle
  \frac{1}{2m_e} \left( \frac{d}{dr} - \frac{\kappa}{r} \right) \pi_{k-n} (r)
  \\[4mm] \displaystyle
  \pi_{k-n} (r)
 \end{array}
  \right), \qquad k>n \, ,
 \label{eq:dkb_basis2}
\end{equation}
where $2n$ is the size of the basis set and $\pi_k$ are linear-independent
functions which are assumed to be square integrable and satisfy the proper 
boundary conditions. In the present work, we have chosen B-splines as
$\pi_k$. The B-splines of any degree can be easily constructed using 
the recursive algorithm~\cite{De_Boor_01, Johnson_88}. With this basis,
the Hamiltonian and overlapping matrices are sparse, which facilitates 
the numerical calculations. 
%
\section{RESULTS}
\label{sec:results}
In this section, we present the results 
of our calculations of the pair-creation probabilities in
the collisions of two identical bare nuclei 
at the energy near the Coulomb barrier. Unless stated otherwise, 
the nuclei are assumed to move along the classical Rutherford trajectories.
The nuclear charge distribution is given by a uniformly charged sphere 
of radius $R_{\rm n}=1.2 \times A^{1/3}$~fm, where $A$ is the atomic mass number. 
The calculations were performed employing the method described in 
Sec.~\ref{sec:theory} for the states with the relativistic quantum number 
$\kappa=-1$ and $\kappa=1$. There is no coupling between these sets 
in the monopole approximation
and they are expected to give the dominant contribution 
to the pair creation~\cite{Reinhardt_81}.
We used 410 basis functions constructed from B-splines of ninth order defined
in a box of size $L=10^5$~fm. The B-spline knots were distributed exponentially in
order to better describe the wave functions in the region of the closest 
approach of the nuclei. It was found that this basis set is sufficient 
to obtain the convergent
results. All of the initial states, including 10 bound, 195 positive-continuum, 
and 205 negative-continuum ones, were 
propagated in order to obtain the one-electron transition amplitudes.
The particle numbers were calculated according to 
the formulas (\ref{eq:electrons}) and (\ref{eq:positrons}) for
a fixed projection $m$ of the total angular momentum $j=1/2$
and were then doubled in order to take into account the contributions of channels
with both values of $m$.

\begin{figure}
\centering 
\includegraphics[trim=0 0 0 0, clip, width = 0.7\textwidth]
{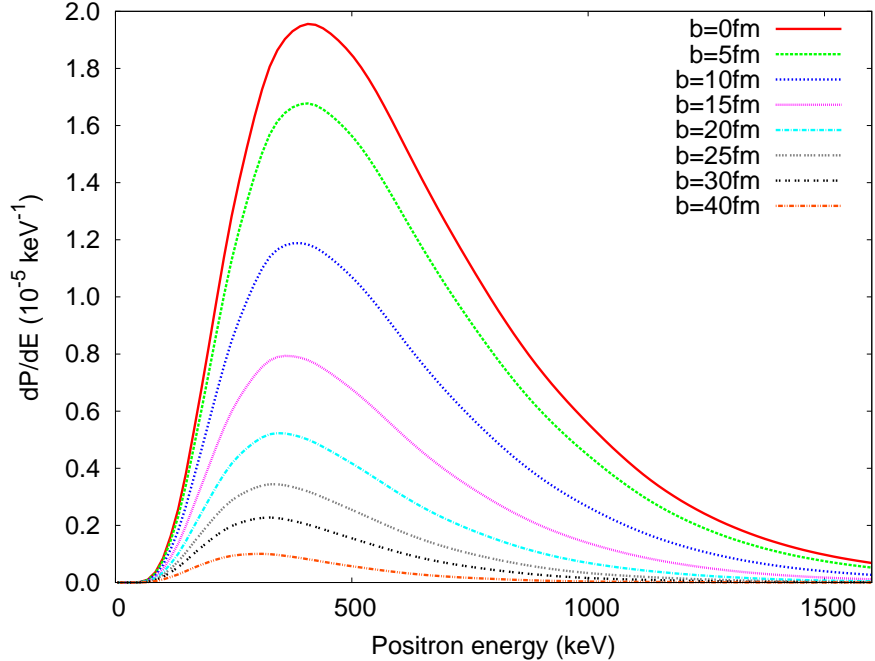}
\caption{(Color online) Positron energy spectrum for the U$-$U collision 
at energy $E_{\rm cm}=$740~MeV for 
the different values of the impact
parameter $b$.
}
\label{fig:u_spectra}
\end{figure}

In Fig.~\ref{fig:u_spectra}, we present the obtained positron energy spectra for the U$-$U
collision for the different values of the impact parameter $b$ at kinetic energy $E_{\rm cm}=$740~MeV 
in the center-of-mass frame. 
These results are very similar to those presented in Ref.~\cite{Muller_88}.
The collisions with $b=30$~fm and $b=40$~fm are subcritical, and 
with $b\leq25$~fm, they are supercritical. However, the calculated positron spectra 
do not exhibit any qualitative difference between the subcritical and supercritical
regimes.

In Table~\ref{tab:pair_numbers}, the obtained numbers of created pairs for 
the U$-$U collision at $E_{\rm cm}=740$~MeV and $E_{\rm cm}=680$~MeV
are presented and compared with the
corresponding values from Ref.~\cite{Muller_88}. The results are in good agreement 
with each other, but in our case the contribution of pairs with a free electron is
relatively larger. This can be due to a more dense representation 
of the continuum states in our calculations. Nevertheless, as one can see from 
Table~\ref{tab:pair_numbers}, the created electrons are mainly captured 
into the bound states.
\begin{table}[ht]
\caption{Number of created pairs in the U$-$U collision at energy $E_{\rm cm}$ 
as a function of the impact parameter $b$. $P_t$ is the total number of pairs
and $P_b$ is the number of pairs with an electron captured into a bound state.}
\label{tab:pair_numbers}
\begin{center}
  \begin{tabular}{|c|c|c|c|c|c|}
  \hline
  \multicolumn{2}{|c|}{} & \multicolumn{2}{|c|}{M\"{u}ller {\it et al.}~\cite{Muller_88}} &
  \multicolumn{2}{|c|}{This work}\\[0mm]
  \hline
  ~$E_{\rm cm}$~(MeV)~&~$b$~(fm)~ & \hspace{7mm}  $P_{\tiny b}$  \hspace{7mm} &
  \hspace{7mm} $P_{t}$ \hspace{7mm} & \hspace{7mm}  $P_{b}$  \hspace{7mm} &
  \hspace{7mm} $P_{t}$  \hspace{7mm} \\[1mm]
  \hline
  740& 0  &   1.23 $\times$ $10^{-2}$  &  ~~1.26 $\times$ $10^{-2}$~~ & 1.25 $\times$ $10^{-2}$ &  1.29 $\times$ $10^{-2}$~~ \\[0mm]
  & 5  &   1.04 $\times$ $10^{-2}$  &    1.06 $\times$ $10^{-2}$   & 1.05 $\times$ $10^{-2}$ &  1.08 $\times$ $10^{-2}$~~ \\[0mm]
  & 10  &  7.04 $\times$ $10^{-3}$  &    7.15 $\times$ $10^{-3}$   & 7.03 $\times$ $10^{-3}$ &  7.26 $\times$ $10^{-3}$~~ \\[0mm]
  & 15  &  4.41 $\times$ $10^{-3}$  &    4.47 $\times$ $10^{-3}$   & 4.39 $\times$ $10^{-3}$ &  4.51 $\times$ $10^{-3}$~~ \\[0mm]
  & 20  &  2.71 $\times$ $10^{-3}$  &    2.73 $\times$ $10^{-3}$   & 2.70 $\times$ $10^{-3}$ &  2.75 $\times$ $10^{-3}$~~ \\[0mm]
  & 25  &  1.67 $\times$ $10^{-3}$  &    1.68 $\times$ $10^{-3}$   & 1.66 $\times$ $10^{-3}$ &  1.69 $\times$ $10^{-3}$~~ \\[0mm]
  & 30  &  1.04 $\times$ $10^{-3}$  &    1.04 $\times$ $10^{-3}$   & 1.03 $\times$ $10^{-3}$ &  1.04 $\times$ $10^{-3}$~~ \\[0mm]
  & 40  &  4.11 $\times$ $10^{-4}$  &    4.11 $\times$ $10^{-4}$   & 4.09 $\times$ $10^{-4}$ &  4.12 $\times$ $10^{-4}$~~ \\[0mm]
  \hline
  680&0  &   1.04 $\times$ $10^{-2}$  &  ~~1.06 $\times$ $10^{-2}$~~ & 1.05 $\times$ $10^{-2}$ &  1.07 $\times$ $10^{-2}$~~ \\[0mm]
  &5  &   8.86 $\times$ $10^{-3}$  &    8.97 $\times$ $10^{-3}$   & 8.87 $\times$ $10^{-3}$ &  9.10 $\times$ $10^{-3}$~~ \\[0mm]
  &10  &  6.05 $\times$ $10^{-3}$  &    6.12 $\times$ $10^{-3}$   & 6.03 $\times$ $10^{-3}$ &  6.17 $\times$ $10^{-3}$~~ \\[0mm]
  &15  &  3.80 $\times$ $10^{-3}$  &    3.83 $\times$ $10^{-3}$   & 3.78 $\times$ $10^{-3}$ &  3.85 $\times$ $10^{-3}$~~ \\[0mm]
  &20  &  2.33 $\times$ $10^{-3}$  &    2.34 $\times$ $10^{-3}$   & 2.32 $\times$ $10^{-3}$ &  2.35 $\times$ $10^{-3}$~~ \\[0mm]
  &25  &  1.43 $\times$ $10^{-3}$  &    1.43 $\times$ $10^{-3}$   & 1.42 $\times$ $10^{-3}$ &  1.44 $\times$ $10^{-3}$~~ \\[0mm]
  &30  &  8.80 $\times$ $10^{-4}$  &    8.80 $\times$ $10^{-4}$   & 8.75 $\times$ $10^{-4}$ &  8.82 $\times$ $10^{-4}$~~ \\[0mm]
  &40  &  3.42 $\times$ $10^{-4}$  &    3.42 $\times$ $10^{-4}$   & 3.41 $\times$ $10^{-4}$ &  3.43 $\times$ $10^{-4}$~~ \\[0mm]
  \hline
  \end{tabular}
\end{center}
\end{table}

\begin{figure}
\centering 
\includegraphics[trim=0 0 0 0, clip, width = 0.7\textwidth]
{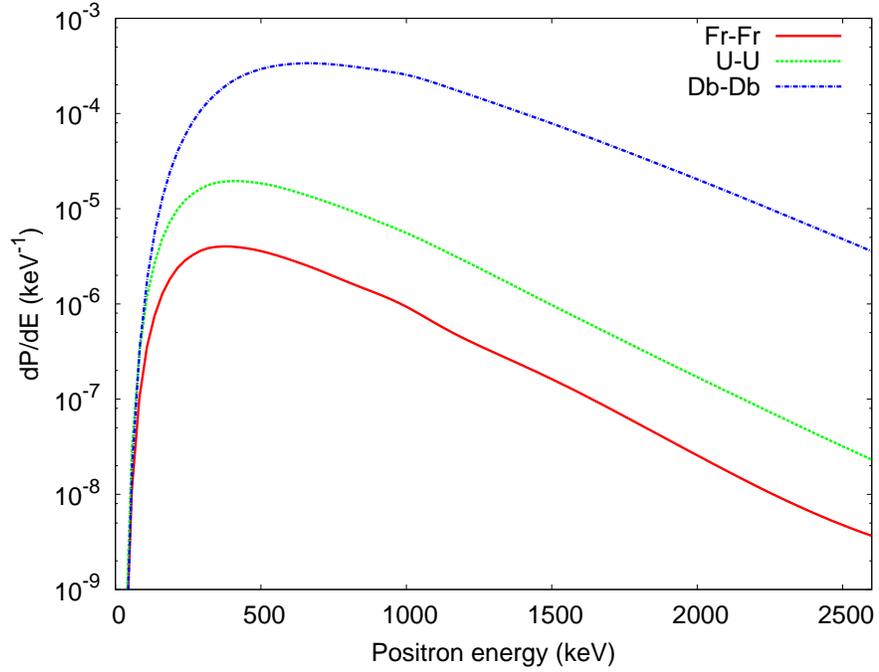}
\caption{(Color online) Positron energy spectrum for the Fr$-$Fr, U$-$U, and Db$-$Db 
head-on collisions at energies 
674.5, 740, and 928.4~MeV, respectively.
}
\label{fig:FrDbU}
\end{figure}

In order to study possible evidences of the spontaneous pair creation, we
considered the collisions of nuclei with different charge $Z$. Figure~\ref{fig:FrDbU} 
shows the obtained positron spectra for the Fr$-$Fr~($Z$=87), U$-$U~($Z$=92), 
and Db$-$Db~($Z$=105) head-on collisions at 
$E_{\rm cm}=674.5$, $E_{\rm cm}=740$, and $E_{\rm cm}=928.4$~MeV, respectively.
For these energies, the minimal distance between the nuclear surfaces is the same 
for all three cases (about $1.6$~fm). The Fr$-$Fr collision is subcritical and has
the purely dynamical positron spectrum. In the Db$-$Db collision, one can expect an
enhancement of the spontaneous pair creation due to the deep supercritical resonance~\cite{Ackad_08}.
However, all three curves in Fig.~\ref{fig:FrDbU} have a similar shape. 
The obtained positron spectra are quite different from those in Ref.~\cite{Ackad_08}, especially 
for the small positron energies. 

\begin{figure}
\centering 
\includegraphics[trim=0 0 0 0, clip, width = 0.7\textwidth]
{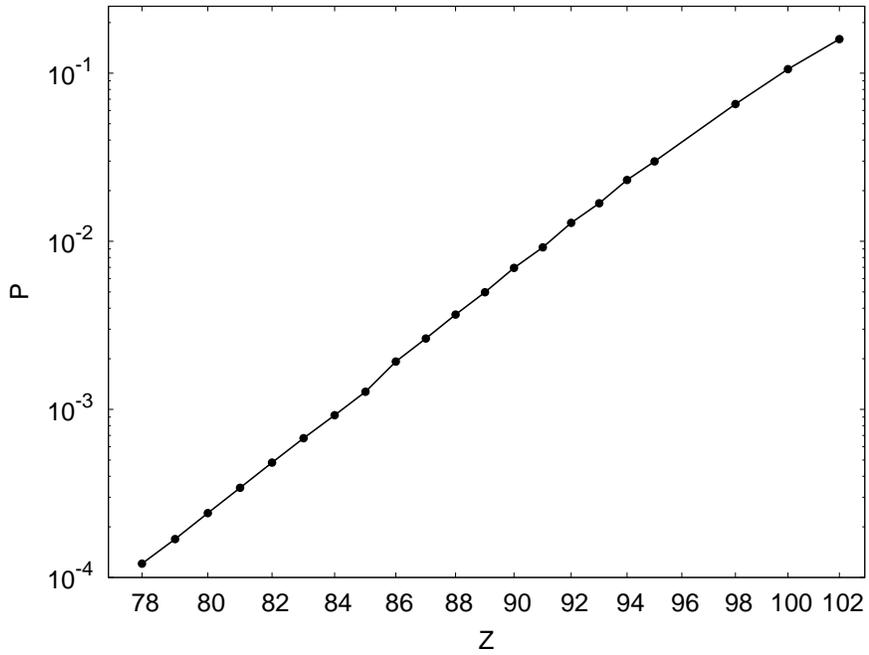}
\caption{Number of created pairs $P$ in the head-on collision of identical
nuclei as a function of the nuclear charge $Z_A=Z_B=Z$
for the projectile energy $E_0=6.2$~MeV/u in the 
nuclear rest frame.}
\label{fig:z_dep}
\end{figure}

In Fig.~\ref{fig:z_dep}, we present the number of created
pairs $P$ in head-on collisions of two identical nuclei as a function 
of the nuclear charge $Z_A=Z_B=Z$ for the projectile energy $E_0=6.2$~MeV/u in the 
nuclear rest frame, which corresponds to $E_{\rm cm}=740$~MeV for the U$-$U collisions. 
There is a very strong dependence of $P$ on $Z$, which in 
the subcritical region $78 \leq Z \leq 87$ can be
parametrized by $Z^\gamma$ with $\gamma \approx 28$. The function
$P(Z)$ smoothly continues into the supercritical region $Z>87$,
but its growth is slowing down for the higher $Z$. This result is very close to the 
corresponding one in Ref.~\cite{Reinhardt_81}, where it was found that in collisions
of bare nuclei, the positron production is proportional to 
$(Z_A+Z_B)^\gamma$ with $\gamma \approx 29$.

\begin{figure}
\centering 
\includegraphics[trim=0 0 0 0, clip, width = 0.7\textwidth]{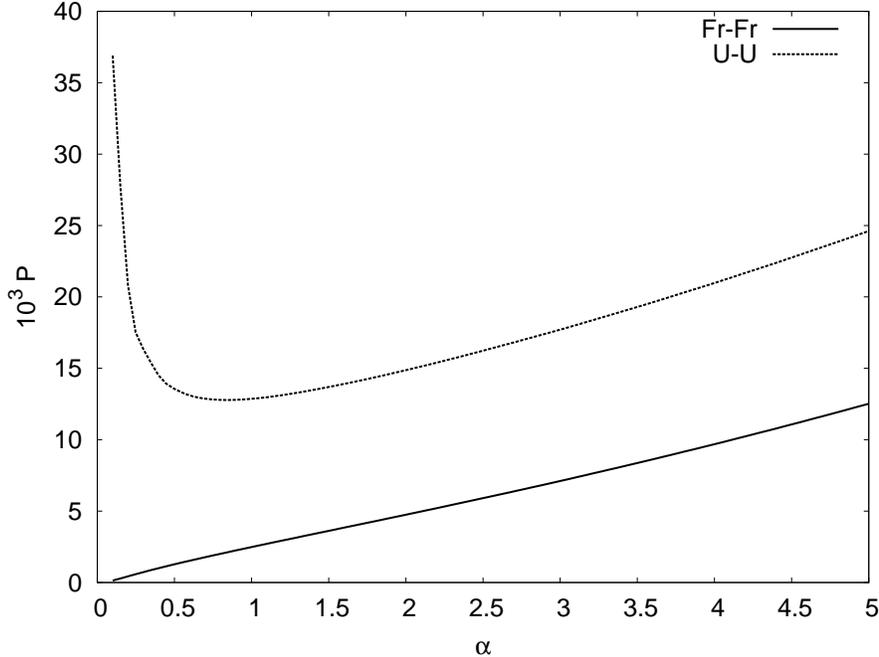}
\caption{Number of created pairs $P$ in the head-on collision with the
artificial trajectory $R_\alpha (t)$, defined by Eq.~(\ref{eq:art_trajectory}), 
as a function of $\alpha$. The solid line indicates
the results for the Fr$-$Fr collision at $E_{\rm cm}=674.5$~MeV; the dashed line corresponds
to the U$-$U collision at $E_{\rm cm}=740$~MeV.}
\label{fig:var_speed}
\end{figure}

In order to demonstrate the ability of our method to describe 
the spontaneous pair creation we considered the supercritical U$-$U and subcritical Fr$-$Fr 
collisions with artificial trajectories at 
$E_{\rm cm}=674.5$ and $E_{\rm cm}=740$~MeV, respectively. 
First, we introduce the new trajectory $R_\alpha (t)$,
\begin{equation}
\dot{R}_\alpha (t)=\alpha \dot{R}(t),
\label{eq:art_trajectory}
\end{equation}
where $R(t)$ is the 
classical Rutherford trajectory. In Fig.~\ref{fig:var_speed}, we present the number of 
created pairs $P$ as a function of $\alpha$ for the U$-$U and Fr$-$Fr head-on collisions.
It can be seen that in both cases $P(\alpha)$ grows monotonically for large $\alpha$, which
can be explained by an enhancement of the dynamical pair production due to the fast variation of 
the potential. For small values of $\alpha$, where the dynamical mechanism is suppressed, 
$P (\alpha)$ increases for the U$-$U collision and goes to zero for the Fr$-$Fr collision, 
which indicates the existence of the spontaneous pair creation in the supercritical case. 

\begin{figure}
\centering 
\includegraphics[trim=0 0 0 0, clip, width = 0.7\textwidth]
{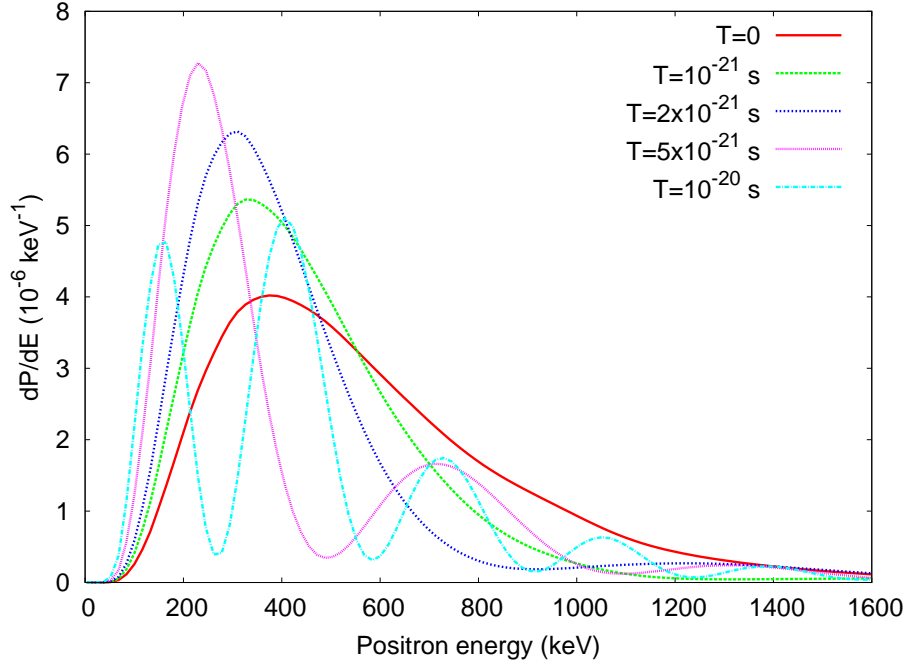}
\caption{(Color online) Positron energy spectrum for the Fr$-$Fr head-on collision 
 at $E_{\rm cm}=674.5$~MeV
 with different time delays $T$.}
\label{fig:fr_stick}
\end{figure}

\begin{figure}
\centering 
\includegraphics[trim=0 0 0 0, clip, width = 0.7\textwidth]
{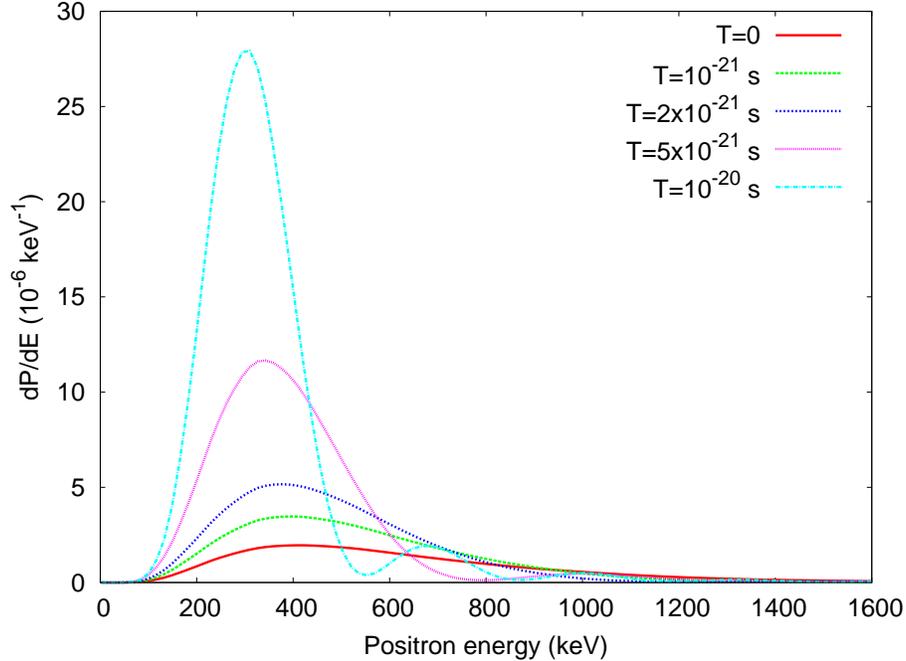}
\caption{(Color online) Positron energy spectrum for the U$-$U head-on collision 
 at $E_{\rm cm}=740$~MeV
 with different time delays $T$.}
\label{fig:u_stick}
\end{figure}

We also considered the trajectories with the time delay $T$ at the closest 
approach of the nuclei. Such trajectories can be used to model 
the hypothetical collisions with the nuclear sticking~\cite{Greiner_85}. In the 
supercritical case the time delay should enhance the spontaneous pair creation.
The obtained positron spectra for the pure Rutherford trajectory ($T=0$) and for
the different time delays in the head-on Fr$-$Fr and U$-$U collisions 
are shown in Figs.~\ref{fig:fr_stick} 
and~\ref{fig:u_stick}, respectively. As can be seen from 
Fig.~\ref{fig:fr_stick}, the shape of the positron spectrum is changed significantly 
with growing $T$. However, the variations of the total number of created pairs $P$ for 
the Fr$-$Fr collisions are less than 15\% and are oscillating. 
In the supercritical U$-$U collisions, $P$ increases monotonically as $T$ grows, 
which demonstrates the enhancement of the spontaneous pair creation. 
It can be seen from the figures that some additional peaks appear for large $T$ in both cases. 
However, in the supercritical case, the main peak is much higher 
than the others and steadily grows and shifts towards the low energies with increasing $T$. 
This leads to the conclusion that the spontaneous mechanism predominantly 
contributes to the region of the main peak for the largest $T$.
Our results for the positron spectra in the U$-$U collisions with the time delay are in good agreement 
with the corresponding ones from Ref.~\cite{Muller_88}, and differ 
from the values obtained in Ref.~\cite{Ackad_08}, especially for the small positron
energies.
%
\section{CONCLUSION}
We presented a method for calculations of pair production
in low-energy collisions of bare nuclei. Using this method, 
the energy spectra of emitted positrons and the numbers of created 
pairs in collisions of identical nuclei 
were calculated in the monopole approximation for different values of the 
impact parameter and the nuclear charge. 
The ability of the method to describe the spontaneous pair creation 
was demonstrated by calculations for the collisions with the modified velocity 
and with the time delays. 

The obtained results for the U$-$U collisions are in good agreement with 
the corresponding values from Ref.~\cite{Muller_88} for all considered
impact parameters. The calculations showed a very strong dependence of 
the dynamical pair creation on the nuclear charge, which confirms the results
of Ref.~\cite{Reinhardt_81}.
The calculated positron energy spectra for the U$-$U, Fr$-$Fr, 
and Db$-$Db head-on collisions disagree with those presented in Ref.~\cite{Ackad_08}. 
The reason for this discrepancy is unclear to us.
%

 A comparison of the different subcritical and supercritical scenarios leads to the conclusion
that no direct evidence of the spontaneous pair creation can be found in
the positron energy spectra for the heavy-ion collisions with 
the Rutherford trajectory.  We expect, however, that the detailed studies
of various processes that take place in low-energy heavy-ion collisions,
including the angular-resolved positron energy spectra, can examine the validity
of QED at the supercritical regime. For these studies, more elaborated
full three-dimensional methods are needed. To date, such methods have
been developed for calculations of the electron-excitation and 
charge-transfer probabilities only~\cite{Tup_10, Tup_12, Kozhedub_2013,
Maltsev_2013, Deyneka_13, Kozhedub_14}. The extension of these
methods to calculations of pair-production probabilities is one of
the main goals of our future work.
\section*{\large Acknowledgments}
We thank Marko Horbatsch and Edward Ackad for valuable discussions.
This work was supported by RFBR (Grants No. 13-02-00630 and No. 14-02-31418), 
by SPbSU (Grants No. 11.38.269.2014 and No. 11.38.654.2013),
and by the President of 
the Russian Federation (Grant No. MK-6970.2015.2).
I.A.M. and Y.S.K. acknowledge the financial support of FAIR-Russia Research Center.
The work of I.A.M. was also supported by the Dynasty foundation and 
by the German-Russian Interdisciplinary Science Center (G-RISC) 
funded by the German Federal Foreign Office via the German Academic 
Exchange Service (DAAD).
%
%
\clearpage


%
\end {document}